\documentclass[submission,copyright,creativecommons]{eptcs}
\providecommand{\event}{ACL2 2020} % Name of the event you are submitting to
\usepackage{breakurl}             % Not needed if you use pdflatex only.
\usepackage{underscore}           % Only needed if you use pdflatex.

\title{RP-Rewriter: An Optimized Rewriter for Large Terms in ACL2}

\author{Mertcan Temel
  \institute{Department of Electrical and Computer Engineering\\
    University of Texas at Austin\\
    Austin, Texas, USA} \email{mert@utexas.edu}
}

\newcommand*{\pcr}{\fontfamily{pcr}\selectfont}

\usepackage{amssymb}

\usepackage{amsthm}

\newtheorem{example}{Example}%[lemma]
\usepackage{pgfplots}

\usepackage{subcaption}

\pgfplotsset{width=7cm,compat=1.9}

\usepackage{verbatimbox}

\usepackage{booktabs,tabularx, colortbl}

\def\titlerunning{RP-Rewriter}
\def\authorrunning{M. Temel}
\begin{document}
\maketitle

\begin{abstract}
  RP-Rewriter ({\it  Retain-Property}) is a verified  clause processor that
  can use  some of  the existing  ACL2 rewrite  rules to  prove conjectures
  through term  rewriting. Optimized for  conjectures that can  expand into
  large  terms, the  rewriter tries  to mimic  some of  the ACL2  rewriting
  heuristics  but   also  adds   some  extra   features.   It   can  attach
  side-conditions to terms  that help the rewriter  retain properties about
  them and prevent possibly some very expensive backchaining.  The rewriter
  supports  user-defined  complex meta  rules  that  can return  a  special
  structure to prevent redundant rewriting. Additionally, it can store fast
  alists even when values are not  quoted.  RP-Rewriter is utilized for two
  applications,  multiplier design  proofs and  SVEX simplification,  which
  involve very large terms.
\end{abstract}

\section{Introduction}

During the  development process of  a new  proof method for  correctness of
multiplier  designs, which  is based  on term-rewriting,  we have  stressed
ACL2's built-in rewriting system for very large terms. Even though it is an
intricate  program that  can prove  various conjectures  with a  variety of
rules, we have found that some  conjectures that can expand into very large
terms  such as  multiplier designs  may pose  a challenge  to the  built-in
rewriter.   In order  to optimize  the performance,  we came  up with  some
additional rewriter features including support for side-conditions (defined
below) that  can prevent  exhaustive backchaining,  a mechanism  to prevent
redundant rewriting of  terms returned from meta functions,  and an ability
for rewrite rules to create fast-alists when keys are quoted and values are
not.   Throughout  the paper,  we  refer  to ACL2+Books  Documentation  for
certain topics with the note ``see  :DOC doc-name'', and they are available
online \cite{doc}.

We introduce a rewriting mechanism  where terms can have certain properties
attached   as  side-conditions   in  order   to  help   relieve  hypotheses
efficiently.   For example,  {\pcr  integerp}  can be  attached  to a  term
indicating that  it satisfies  {\pcr integerp}.  Retaining  such properties
about   terms  can   provide   some  performance   benefits   as  well   as
convenience. For example, consider the  basic {\it logop} (see :DOC logops)
and {\it  4vec} (see :DOC  sv::4vec) functions.  The {\it  log} operations,
such  as  {\pcr  logapp},  {\pcr  loghead} and  {\pcr  logand},  work  with
integers,  which  may  represent  two-valued  bit  vectors,  for  bit-level
operations.  Most  of the  logops have a  corresponding 4vec  function that
performs  the same  operation  for an  extended  domain (i.e.,  four-valued
bit-vectors).   Assume  that  we  have  a library  of  rewrite  rules  that
efficiently simplifies large terms composed  of 4vec functions.  If we want
to apply the  same simplification algorithm for a term  composed of logops,
then  we could  either  copy and  prove the  same  rules for  corresponding
logops, or we could more easily  rewrite logops to their corresponding 4vec
functions.  The latter is a more practical approach; however, as we rewrite
logops to 4vec  functions, we lose the information that  the rewritten term
satisfies {\pcr integerp}, granted the  logops in question always return an
integer. With  the built-in  rewriter, if  we are  working with  very large
terms, then we would have to backchain  to the innermost terms to show that
the  term satisfies  {\pcr  integerp}.   However, if  we  can attach  {\pcr
  integerp} as a  side-condition to logop terms as we  rewrite them to 4vec
functions, then we can preserve the  property of the term that it satisfies
{\pcr integerp} no  matter what it or its subterms  become after rewriting.
Such a system can  help create a unitary library of  rewrite rules for both
function families with a lower maintenance cost.

Regular meta rules in ACL2 may  cause redundant rewriting of terms that may
entail  performance issues  when  the  terms in  question  are very  large.
ACL2's built-in rewriter works in an inside-out manner when reasoning about
conjectures. Whenever a rewrite rule is applied, the system is smart enough
not to  rewrite the already processed  inner terms. It can  accomplish that
because rewrite rules  have a well defined pattern; and  it is possible for
the  rewriter to  tell apart  the already  processed terms  from the  newly
introduced terms. However, when a meta rule is applied, the associated meta
function  can change  the term  in any  way, which  is not  visible to  the
rewriter.  Therefore,  terms  returned  from  a  meta  rule  are  rewritten
completely as  if everything is  new. In cases  where meta rules  work with
large terms, this redundancy may cause some performance issues.

For terms that include {\pcr  hons-acons} and {\pcr hons-get}, the built-in
rewriter uses logical definitions when at least one argument is not quoted.
However, it  is possible to develop  a system that can  create a fast-alist
whose keys  are quoted  and values  are not.   For conjectures  that create
fast-alists, having such an ability may improve proof-time performance.

With  the goal  of addressing  the aforementioned  three problems,  we have
developed a customized rewriter, RP-Rewriter ({\it Retain-Property}), which
we have published in the  ACL2 Community books (books/projects/rp-rewriter/) as a
verified         clause         processor.         RP-Rewriter         uses
meta-extract~\cite{meta-extract}  to  retrieve  and  utilize  a  subset  of
existing ACL2 rewrite rules. Rewriting is  done in a fashion similar to the
built-in rewriter  with much simpler  heuristics, and we  add user-friendly
features      to      overcome      these     three      problems.       In
Section~\ref{sec:implementation}, we outline the steps taken when rewriting
occurs,  and we  discuss the  implementation details  that set  RP-Rewriter
apart from other rewriters.  In Section~\ref{sec:uses}, we briefly describe
two     applications     where     RP-Rewriter     is     utilized.      In
Section~\ref{sec:support},  we   state  the  supported  rewrite   rules  by
RP-Rewriter.  Finally,  we briefly  discuss some  important details  of the
verification of RP-Rewriter in Section~\ref{sec:proof}.

\section{Implementation}
\label{sec:implementation}

% In this section, we describe implementation details and verification
% methodology for rp-rewriter.

When  the clause  processor function  for RP-Rewriter  is called,  it first
parses   and   gathers   designated  rewrite   rules   using   meta-extract
\cite{meta-extract}. Then the conjecture and parsed rules are passed to our
main rewriter function, {\pcr rp-rw}.  Upon calling {\pcr rp-rw}, we follow
the steps below:
\begin{itemize}
\item Check if rewriting should  terminate with {\it dont-rw} (described in
  the subsection below).
\item If the rewriter is in  if-and-only-if context, try to reduce the term
  to {\pcr 't} by checking existing side-conditions.
\item When  applicable, update the term  with the current context  (i.e., a
  list  of  known  facts  that  may  come  from  hypotheses  or  {\pcr  if}
  expressions).
\item If the  term is of the  form {\pcr (if test then  else)}, rewrite the
  arguments accordingly by  expanding the context; in  other cases, rewrite
  the arguments (i.e., subterms) by making recursive calls to {\pcr rp-rw}.
\item Try  applying executable-counterpart.  For  the current term,  if the
  arguments are  quoted and the  executable counterpart of the  function is
  enabled,   then   run   the   function   using   {\pcr   magic-ev-fncall}
  \cite{meta-extract}.
\item Try  applying meta rules.  If the term  is changed,  recursively call
  {\pcr rp-rw} on the resulting term.
\item Try applying rewrite rules. If  the term is changed, recursively call
  {\pcr rp-rw}  on the resulting  term; otherwise, terminate  rewriting and
  return the current term.
\end{itemize}
\noindent
These  steps  define the  outline  of  our  rewriting heuristics.   In  the
subsections  below, we  describe  the features  and  implementation of  our
rewriter that  sets it apart from  most rewriters.  We start  with the {\it
  dont-rw} structure  that determines  what terms  should be  rewritten and
when  rewriting should  terminate.   Secondly, we  describe  the notion  of
side-conditions with a few  examples to show how it can be  used to prove a
conjecture when the  built-in rewriter struggles, and how it  can be faster
than  the built-in  rewriter. Then  we  briefly talk  about the  fast-alist
support as well as the meta functions that can be used with RP-Rewriter.

\subsection{Controlled Rewriting with ``dont-rw'' Structure}
\label{sec:dont-rw}

We use a special data structure, called  {\it dont-rw}, that we pass to our
rewriter, {\pcr rp-rw}, along with the  term to rewrite in order to control
rewriting and terminate when necessary. We traverse this data structure the
same way as the term being rewritten; that is, we call {\pcr car} and {\pcr
  cdr}  on the  term and  {\it dont-rw}  at the  same time.   Whenever {\it
  dont-rw}  is an  atom and  non-nil, the  rewriter stops  and returns  the
current term.

\begin{example} An  example term being  rewritten and a  corresponding {\it
    dont-rw}.  None  of  the  f3  terms will  be  rewritten  because  their
  corresponding entry  in {\it dont-rw} is  a non-nil atom (i.e.,  {\pcr y}
  and {\pcr z}).

  {\normalfont \pcr
    (f1 (f2 a (f3 b c)) (f4 (f3 b c))) }

  {\normalfont \pcr
    (f1 (f2 x y) (f4 z)) }
\end{example}

We update  {\it dont-rw} whenever a  rule is applied and  the term changes.
When an  applicable rewrite rule, which  has the form {\pcr  (implies\ hyp\
  (equal\ lhs\  rhs))}, is  found, we  unify the term  with {\pcr  lhs}. We
create a new term by applying  the bindings from unification to {\pcr rhs};
and we  pass the rule's {\pcr  rhs} as {\it  dont-rw}.  We do the  same for
{\pcr hyp} when  reliving hypotheses. We give meta functions  the option to
return {\it dont-rw} so that they can attain control over what terms should
be rewritten. When  meta functions do not return {\it  dont-rw}, then it is
set  to {\pcr  nil}  and everything  is rewritten  by  default.  This  {\it
  dont-rw} structure  is invisible to  the users  expect for when  they are
creating meta functions.

\subsection{Side-conditions}
\label{sec:side-cond}

We develop  a system to  attach properties (side-conditions) to  terms that
may be useful  while relieving hypotheses.  For that purpose,  we define an
identity function {\pcr rp} with signature {\pcr (rp prop term)}, where the
first argument {\pcr  prop} is ignored and the second  argument {\pcr term}
is returned unchanged.  When a term  has a side-condition, it is wrapped by
this {\pcr rp} function with its  property to be retained. This property is
a  quoted function  symbol  with  an arity  of  1  (e.g., {\pcr  'integerp}
implying that the term is an integer).  We maintain an invariant during the
rewriting process  that the attached  side-condition is correct.   Below we
describe  how side-conditions  for  terms  are created  with  a few  simple
examples;  and   another  example   showing  how  side-conditions   can  be
instrumental in proving some conjectures.

Terms might attain side-conditions through  meta rules or rewrite rules. In
meta  functions,  users   might  simply  create  an   {\pcr  rp}  instance,
correctness of which users would have  to prove. This may be challenging in
some cases and it is tailored more for advanced users. A more user-friendly
to attach side-conditions can be done through rewrite rules. After deciding
the properties  to retain for certain  terms on the right  hand side, users
can simply prove  a lemma for those side-conditions; and  attach that lemma
to the rewrite rule with a simple  utility. That way, whenever such a rewrite
rule  is   applied,  designated   terms  can   soundly  retain   the  given
side-conditions.

Example \ref{side-cond-example} shows how  a side-condition can be attached
to a rewrite rule. When we  rewrite {\pcr logand} to {\pcr 4vec-bitand}, we
lose  the immediate  knowledge that  the rewritten  term can  satisfy {\pcr
  integerp}. After  rewriting {\pcr  logand} terms with  the first  rule in
this  example, we  would have  to  use the  second  rule to  show that  the
resulting {\pcr 4vec-bitand} term  satisfies {\pcr integerp}, and backchain
(again),  which  can  be  very   expensive  if  arguments  are  very  large
terms. Instead,  we attach this property  as a side-condition to  the first
rule  with {\pcr  rp-attach-sc}.  When  RP-Rewriter starts  to process  the
rules, it merges the two rules into  one, and replaces all the instances of
{\pcr (4vec-bitand x  y)} with {\pcr (rp 'integerp (4vec-bitand  x y))}. If
the  rewriter later  runs  into a  term  of the  form  {\pcr (integerp  (rp
  'integerp  ...)}, then  this  term  is replaced  with  {\pcr  't} if  the
conjecture is in if-and-only-if context.  It is required that the conjuncts
from the hypotheses of the side-condition  lemmas are identical or a subset
of the  conjuncts from the  hypotheses of the  main rule.  Note  that {\pcr
  def-rp-rule} is a macro that first calls {\pcr defthm} and then saves the
rule to be used by RP-Rewriter.  % Example {\ref
%   side-cond-example-2} shows how side-conditions help prevent
% backchaining when a term composed of {\pcr logand} is rewritten with
% {\pcr logand-to-4vec-bitand}.

\begin{example}
  \label{side-cond-example}
  An example of how a side-condition can be attached to
  a rule.
\begin{verbnobox}[\fontfamily{pcr}\selectfont\fontsize{9}{12}\selectfont]
  (def-rp-rule logand-to-4vec-bitand
    (implies (and (integerp x)
                  (integerp y))
             (equal (logand x y)
                    (4vec-bitand x y))))

  (defthm logand-to-4vec-bitand-side-cond
    (implies (and (integerp x)
                  (integerp y))
             (integerp (4vec-bitand x y))))

  (rp-attach-sc logand-to-4vec-bitand
                logand-to-4vec-bitand-side-cond)

\end{verbnobox}
\end{example}

The process  of attaching  side-conditions is  invisible to  users.  During
unification, terms inside of {\pcr rp}  functions are extracted in order to
make side-conditions invisible to the unification process. Therefore, users
do not  need to  consider side-conditions when  creating rewrite  rules; in
fact, users  are not allowed to  have an {\pcr  rp} call in a  rewrite rule
supplied  to   {\pcr  def-rp-rule}.  Such   rules  would  be   rejected  by
RP-Rewriter.  The only times users need  to be aware of this side-condition
mechanism are when writing meta functions and using {\pcr syntaxp}.

Side-conditions may be  useful in proving conjectures that  can expand into
very large  terms. For example,  as discussed  in the introduction,  we may
want to  utilize a simplification  library for 4vec functions  when working
with logops.  We can simply rewrite  all the logops to  their corresponding
4vec  functions  while  remembering  that  all  such  terms  satisfy  {\pcr
  integerp}.  This can  help with  the performance  when trying  to relieve
{\pcr  integerp} of  very large  terms.  Example  \ref{side-cond-example-2}
shows  a  term that  is  rewritten  with {\it  logand-to-4vec-bitand}  from
Example  \ref{side-cond-example}.   In  this  example,   every  4vec-bitand
instance is known to be an integer; and  if we need to prove that this term
is an integer, we can do that without any backchaining. When such terms are
very large, this can reduce memory use and improve proof-time performance.

\begin{example}
  \label{side-cond-example-2}
  Two terms before and after we rewrite with logand-to-4vec-bitand
  from Example \ref{side-cond-example}. 
\begin{verbnobox}[\fontfamily{pcr}\selectfont\fontsize{9}{12}\selectfont]
  (logand (logand x y)
          (logand a b))
\end{verbnobox}        
\begin{verbnobox}[\fontfamily{pcr}\selectfont\fontsize{9}{12}\selectfont]
  (rp 'integerp
      (4vec-bitand (rp 'integerp (4vec-bitand x y))
                   (rp 'integerp (4vec-bitand a b))))
\end{verbnobox}
\end{example}

\begin{figure}
  \centering
  \begin{subfigure}{.47\textwidth}
    \centering
    \begin{tikzpicture}
      \begin{axis}[
        axis lines = left,  
        legend pos=north west,
        xlabel = Depth,
        ylabel = {Runtime (seconds)},
        ymajorgrids=true,
        xmax=18.5,
        ymax=95,
        grid style=dashed,        
        ]
        % Below the red parabola is defined
        % Here the blue parabloa is defined
        \addplot [
        domain=-0:20, 
        samples=100, 
        color=black,
        mark=triangle,
        ]
        coordinates {
          % (5,0.01)
          (8,0.04) (10, 0.17) (12, 0.77) (13, 1.66) (14, 3.70) (15, 8.01) 
          (16, 18.23) (17, 38.68) (18, 89.83)

          % (500,0.04) (2000, 0.17) (8000, 0.77) (16000, 1.66) (32000, 3.70) (64000, 8.01) 
          % (128000, 18.23) (256000, 38.68) (512000, 89.83)
          
          % (20, 23.91)
          % (15, 23592.960) (16, 47185.920) (18, 188743.680) (20, 754974.720)
          % (5, 32.768) (8, 163.840) (10, 720.869)(12, 2949.120) (14, 11796.120)
        };
        \addlegendentry{ACL2 Rewriter}
        \addplot [
        domain=-0:20, 
        samples=100, 
        color=black,
        mark=dash,
        ]
        coordinates {
          % (5,0.04)
          (8,0.05) (10, 0.07) (12, 0.13) (13, 0.16) (14, 0.37) (15, 0.73) 
          (16, 1.40) (17, 2.41) (18, 5.70) % (19, 9.96) %(20, 23.91)  
          % (15, 23592.960) (16, 47185.920) (18, 188743.680) (20, 754974.720)
          % (5, 32.768) (8, 163.840) (10, 720.869)(12, 2949.120) (14, 11796.120)
        };
        \addlegendentry{RP-Rewriter\ \ \ \ \ }
      \end{axis}
    \end{tikzpicture}
    %\caption{}  
  \end{subfigure}  \hskip 15pt
  \begin{subfigure}{.47\textwidth}
    \centering
    \begin{tikzpicture}
      \begin{axis}[
        axis lines = left,  
        legend pos=north west,
        xlabel = Depth,
        ylabel = {Allocated Memory (GB)},
        ymajorgrids=true,
        grid style=dashed,
        xmax=18.5,
        ymax=9.5,
        ]
        % Below the red parabola is defined
        % Here the blue parabloa is defined
        \addplot [
        domain=-0:20, 
        samples=100, 
        color=black,
        mark=triangle,
        ]
        coordinates {
          % (5,0.01)
          (8, .004751) (10, .021462) (12, .098564) (13, .210295) (14, .447062) (15, .946688) 
          (16, 1.999398) (17, 4.211876) (18, 8.846585) %(20, 23.91)
          % (15, 23592.960) (16, 47185.920) (18, 188743.680) (20, 754974.720)
          % (5, 32.768) (8, 163.840) (10, 720.869)(12, 2949.120) (14, 11796.120)
        };
        \addlegendentry{ACL2 Rewriter}

        \addplot [
        domain=-0:20, 
        samples=1000, 
        color=black,
        mark=dash,
        ]
        coordinates {
          % (5,0.04)
          (8, .003370) (10, .006092) (12, .017360) (13, .032561) (14, .063093) (15, .124188) 
          (16, .247343) (17, .495856) (18, .992884) % (19, 9.96) %(20, 23.91)  
          % (15, 23592.960) (16, 47185.920) (18, 188743.680) (20, 754974.720)
          % (5, 32.768) (8, 163.840) (10, 720.869)(12, 2949.120) (14, 11796.120)
        };
        \addlegendentry{RP-Rewriter\ \ \ \ \ }
      \end{axis}
    \end{tikzpicture}
    %\caption{}  
  \end{subfigure}
  
  \caption{Performance comparison of ACL2's built-in rewriter and
    RP-Rewriter on a conjecture with a term tree of 4vec-bitand and
    logapp functions only using the rewrite rules from
    Example~\ref{side-cond-example}. }
  \label{fig:acltwovsrp}
\end{figure}
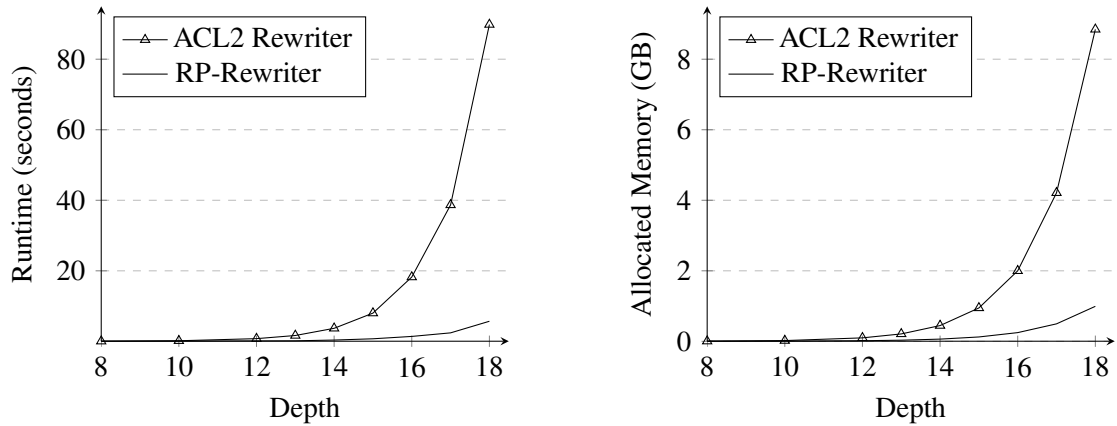

Figure~\ref{fig:acltwovsrp} shows a performance comparison for the built-in
rewriter  and RP-Rewriter  on the  same conjectures,  which highlights  the
performance benefits of side-conditions.  We created dummy conjectures that
have a binary  tree of {\pcr 4vec-bitand} functions on  the left hand side,
and an equivalent binary tree of {\pcr  logand} on the right hand side. For
example, if  the tree depth is  3, then the  left hand size would  have $7$
($1+2+4$) {\pcr  4vec-bitand} instances  nested within  each other  as seen
below:
\begin{verbnobox}[\fontfamily{pcr}\selectfont\fontsize{10}{12}\selectfont]
  (thm
   (equal (4vec-bitand (4vec-bitand (4vec-bitand (iassoc 0 env) ...)
                                    (4vec-bitand (iassoc 2 env) ...))
                       ...)
          (logand ... ...)))
\end{verbnobox} 
\noindent
The innermost elements  are of the form {\pcr (iassoc  x env)}, where {\pcr
  iassoc} is known to return an integer, {\pcr x} is a unique constant, and
{\pcr env} is the only variable  in all of the conjecture.  Before starting
the  proofs,  we  set  the  theory with  {\pcr  minimal-theory}  (see  :DOC
minimal-theory) for both rewriters, and enabled only the rewrite rules from
Example~\ref{side-cond-example} and  {\pcr integerp}  of {\pcr  iassoc}. We
used  {\pcr  thm} and  {\pcr  rp-thm}  (see  :DOC  rp-thm) to  prevent  the
additional cost of saving rules.  We did not provide any hints, or made any
attachments  to the  built-in rewriter.   We gradually  increased the  tree
depth of the conjectures and measured the overall time and allocated memory
of each  event.  As seen  in the figure, the  built-in rewriter has  a much
sharper incline  in resource allocation in  terms of both time  and memory.
This is mainly because the built-in  rewriter has to backchain to innermost
elements every time it needs  to apply {\pcr logand-to-4vec-bitand} whereas
RP-Rewriter uses  the attached side-conditions  and does not  backchain. We
have run  these experiments with SBCL  2.0.2 on a machine  with an Intel(R)
Xeon(R) CPU E1270 @ 3.50GHz.

% Here ends the furst plot
% \hskip 5pt
% % Here begins the 3d plot
% \begin{tikzpicture}
% \begin{axis}
% \addplot[color=red]{exp(x)};
% \end{axis}
% \end{tikzpicture}

There may be other cases where having the side-condition system can provide
much more convenience  than only performance, and it  may prove conjectures
that ACL2's rewriter cannot. Example~\ref{d-evenp-example} shows the events
to create such a case.  With these events (also published in ACL2 Community
Books in projects/rp-rewriter/demo.lsp), we  try to prove some conjectures
such as {\pcr three-round-to-evens}.

\hskip 1 pt

%\vspace {pt}
% \nobreak

\break

\begin{example}
  \label{d-evenp-example}
  A set of events to create a case where RP-Rewriter can prove a
  conjecture with side-conditions when ACL2's built-in rewriter fails.
\begin{verbnobox}[\fontfamily{pcr}\selectfont\fontsize{9}{12}\selectfont]
  (encapsulate
    (((d2 *) => *) ((f2 *) => *) ((neg-m2 *) => *))
    (local (include-book "arithmetic-5/top" :dir :system))

    (local (defun d2 (x) (/ x 2)))
    (local (defun f2 (x) (floor x 2)))
    (local (defun neg-m2 (x) (- (mod x 2))))

    ;; syntaxp is necessary because rp-rewriter does not support loop-stopper.
    (def-rp-rule +-comm
      (implies (syntaxp (and (not (lexorder y x))
                             (or (atom x)
                                 (not (equal (car x) 'binary-+)))))
               (and (equal (+ y x) (+ x y))
                    (equal (+ y x z) (+ x y z)))))

    (def-rp-rule my+-assoc
      (equal (+ (+ a b) c) (+ a b c)))

    (progn
      ;; a lemma for RP-Rewriter to maintain evenness
      (def-rp-rule my+-assoc-for-evens
        (implies (and (evenp (+ a b)) (evenp c))
                 (equal (+ (+ a b) c) (+ a b c))))

      (defthmd my+-assoc-for-evens-side-cond
        (implies (and (evenp (+ a b)) (evenp c))
                 (evenp (+ a b c))))

      (rp-attach-sc my+-assoc-for-evens
                    my+-assoc-for-evens-side-cond))

    (def-rp-rule d2-is-f2-when-even
      (implies (evenp x)
               (equal (d2 x) (f2 x))))

    ;; a function that rounds down a number to an even value
    ;; e.g., (round-to-even 93/10) = 8
    
    (defun round-to-even (a)
      (+ a (neg-m2 a)))

    ;; add definition rule to rp-rewriter's rule-set
    (add-rp-rule round-to-even)

    ;; rhs of the definition rule
    (defthmd round-to-even-is-even
      (evenp (+ a (neg-m2 a))))

    (rp-attach-sc round-to-even
                  round-to-even-is-even))

  (acl2::must-fail
   (defthm three-round-to-evens
     (equal (d2 (+ (round-to-even a) (round-to-even b) (round-to-even c)))
            (f2 (+ (neg-m2 a) (neg-m2 b) (neg-m2 c) a b c)))))
  (acl2::must-succeed
   (defthmrp three-round-to-evens-use-rp ;; use RP-Rewriter as a clause-processor
     (equal (d2 (+ (round-to-even a) (round-to-even b) (round-to-even c)))
            (f2 (+ (neg-m2 a) (neg-m2 b) (neg-m2 c) a b c)))))
  (acl2::must-succeed
   (defthmrp four-round-to-evens ;; use RP-Rewriter as a clause-processor
     (equal (d2 (+ (round-to-even a) (round-to-even b)
                   (round-to-even c) (round-to-even d)))
            (f2 (+ (neg-m2 a) (neg-m2 b) (neg-m2 c) (neg-m2 d) a b c d)))))
\end{verbnobox}
\end{example}

When the  ACL2's built-in rewriter works  on the conjecture given  in {\pcr
  three-round-to-evens},  it   first  expands  the  definitions   of  {\pcr
  round-to-even} instances. Then the left-hand-side becomes:
\begin{verbnobox}[\fontfamily{pcr}\selectfont\fontsize{10}{12}\selectfont]
  (d2 (+ (+ a (neg-m2 a)) (+ b (neg-m2 b)) (+ c (neg-m2 c))))}
\end{verbnobox} 
\noindent
Then using commutativity and associativity of {\pcr +}, this term
becomes:
\begin{verbnobox}[\fontfamily{pcr}\selectfont\fontsize{10}{12}\selectfont]
  (d2 (+ a b c (neg-m2 a) (neg-m2 b) (neg-m2 c))).
\end{verbnobox}
\noindent
The  attempt to  apply {\pcr  d2-is-f2-when-even} fails  because it  cannot
prove that  the argument is {\pcr  evenp}.  The individual groups  of terms
that make  this argument  even are distributed  across the  summation.  For
this specific problem,  users might derive various  solutions; for example,
it is  possible to  prove a  lemma that  terms of  this form  satisfy {\pcr
  evenp},  or  alternatively we  can  set  {\pcr  neg-m2} as  an  invisible
function for {\pcr  +} for the loop-stopper algorithm (this  keeps {\pcr a}
and {\pcr neg-m2} next to each other)  and prove some simple lemmas to show
evenness. However, for  cases where terms are very large  and the arguments
are rewritten  to different  terms, which  is the  case for  our multiplier
design  proofs \cite{mult-paper},  then  neither of  these solutions  would
work. In fact  for our multiplier proofs,  we have not been able  to find a
feasible solution using the built-in rewriter, other than possibly creating
a complex meta rule to show evenness, which would likely be very tedious to
implement and costly for proof-time performance.

On the other hand, with the retained side-conditions, RP-Rewriter can prove
the same conjecture  as seen in {\pcr  three-round-to-evens-use-rp}, where {\pcr
  defthmrp}  is a  macro  that creates  a {\pcr  defthm}  event that  calls
RP-Rewriter as a clause processor. When RP-Rewriter works on the same term,
the left-hand-side becomes:
\begin{verbnobox}[\fontfamily{pcr}\selectfont\fontsize{10}{12}\selectfont]
  (d2 (+ (rp 'evenp (+ a (neg-m2 a)))
         (rp 'evenp (+ b (neg-m2 b)))
         (rp 'evenp (+ c (neg-m2 c)))))
\end{verbnobox} 
\noindent
With the given lemmas, the terms in summations are ordered
while maintaining evenness, and we get:
\begin{verbnobox}[\fontfamily{pcr}\selectfont\fontsize{10}{12}\selectfont]
  (d2 (rp 'evenp (+ a b c (neg-m2 a) (neg-m2 b) (neg-m2 c)))).
\end{verbnobox} 
\noindent
Then we  can apply  {\pcr d2-is-f2-when-even} because  the evenness  of the
argument  is  maintained  and  hypothesis is  relieved  without  using  any
lemma. The same system is sufficient  for other similar conjectures such as
{\pcr four-round-to-evens}. Due  to readability, we cannot show  a case for
large terms  where subterms might  change drastically; however, we  use the
same system for cases where there may be millions of nodes under subterms.

\subsection{Fast-alist Support for Non-quoted Terms}
\label{sec:fast-alist}

For  conjectures  that  expand  into  terms  with  big  association  lists,
RP-Rewriter implements  a system  to support  creation of  fast-alists when
keys are quoted and  values are not. When a rewrite  rule introduces a term
with {\pcr  hons-acons} with  a quoted  key, instead  of using  the logical
definition   of   {\pcr   hons-acons},   we   create   a   {\it   shadowing
  fast-alist}. When  an instance of  {\pcr hons-get} is introduced  for the
same association list, we retrieve entries from the shadowing fast-alist.

The described fast-alist  mechanism is a part of the  verified rewriter. We
create an  identity function {\pcr  falist} with a signature  {\pcr (falist
  fast-alist term)} where the first argument  is ignored and the second one
is returned. The first argument of this function is always quoted and it is
the  shadowing  fast-alist that  corresponds  to  the  term in  the  second
argument.  This  mechanism is  also invisible to  users, and  {\pcr falist}
instances  are  created  and  updated  with  {\pcr  hons-acons}  and  {\pcr
  fast-alist-free}. We maintain an invariant about {\pcr falist} instances,
that is,  the first and second  arguments are association lists  in correct
syntactic shape  and always  have matching entries.   Similar to  {\pcr rp}
instances,  users are  not  allowed to  introduce  {\pcr falist}  instances
through rewrite rules.

Assume that we introduce three {\pcr hons-acons} instances on an empty
alist ({\pcr 'nil}) with three different quoted keys and distinct
values. The complete term is:
\begin{verbnobox}[\fontfamily{pcr}\selectfont\fontsize{10}{12}\selectfont]
  (hons-acons 'key1 val1
              (hons-acons 'key2 val2
                          (hons-acons 'key3 val3 'nil)))  
\end{verbnobox} 
\noindent
RP-Rewriter would create a {\pcr falist} instance and would rewrite
this term to:
\begin{verbnobox}[\fontfamily{pcr}\selectfont\fontsize{10}{12}\selectfont]
  (falist '((key1 . val1) (key2 . val2) (key3 . val3))
          (cons (cons 'key1 val1)
                (cons (cons 'key2 val2)
                      (cons (cons 'key3 val3) 'nil))))  
\end{verbnobox} 
\noindent
This first argument of this term is our shadowing fast-alist, and the
second argument is the logical equivalent of the input term.

This support  for fast-alist can  provide significant benefits  for certain
applications. We  have tested this feature  with an earlier version  of our
multiplier proof library  where we could enable and  disable the fast-alist
feature easily. Our multiplier correctness proofs create a large alist when
expanding  the definition  of multiplier  designs  to store  the values  of
internal  wires. When  the  fast-alist feature  is  enabled, the  described
mechanism  is deployed  and {\pcr  hons-get} retrieves  elements from  that
alist; when it is disabled, another  meta rule parses the term representing
the alist  to retrieve elements.  Table~\ref{fast-alist-table}  shows these
experimental  results for  some  Wallace and  Dadda  tree multipliers  with
simple  partial products  and simple  final stage  adder. As  seen on  this
table, the  effect of our fast-alist  support can be very  significant with
larger circuits. We have run these experiments with SBCL 2.0.2 on a machine
with an Intel(R) Xeon(R) CPU E1270 @ 3.50GHz.

\begin{table}[htbp]
  \caption{Runtime performance of the fast-alist feature on some multiplier
    correctness proofs (seconds)}
  \label{fast-alist-table}
    \centering
    \begin{tabularx}{\textwidth}{
       p{0.17\textwidth}
       p{0.17\textwidth}
       >{\centering\arraybackslash}
       p{0.27\textwidth}
       >{\centering\arraybackslash}
     p{0.27\textwidth}
   } %{| c | c | c | c |}
        \hline
        Size & Multiplier Type & Fast-Alist Feature Enabled & Fast-Alist Feature Disabled \\
        \hline
          32x32   &  Dadda     &  0.58            &  0.75 \\
                  &  Wallace   &  0.80            &  0.92  \\
          64x64   &  Dadda     &  3.24            &  5.75  \\
                  &  Wallace   &  5.24            &  7.82  \\
          128x128 &  Dadda     &  27.33           &  83.26  \\
                  &  Wallace   &  39.05           &  101.03  \\
          \hline
        \end{tabularx}
        
\end{table}

% \begin{table}[]
%  \caption{Proof-time results for various multipliers in seconds}
%  \label{multi-table1}

%  \begin{tabularx}{\textwidth}{| c | c | c | c |}
%    % {\textwidth}{@{}
%    %    p{0.1\textwidth}
%    %   p{0.1\textwidth}
%    %   p{0.2\textwidth}
%    %   p{0.3\textwidth}
%    %    p{0.3\textwidth}
   
%    % } 
%    %{@{}ll*{10}{C}c@{}}

%   \toprule

%   & Size & Multiplier Type & With Fast-Alist Support & Without Fast-Alist Support  \\

%   % \toprule
%   % I/O      & Benchmark       & \cfnote{\cite{Mahzoondac2019}}{i\hspace{1mm}} & \cite{dkaufmann-FMCAD19}  &  Our Tool   \\
%   \midrule

%   64x64   &  sp-dt-bk   &  39          &   \\
%           &  sp-wt-lf   &  33          &   \\
%   128x128 &  sp-cwt-ks  &  33   &   \\
%           &  sp-ar-rc   &  23          &   \\
%   256x256 &  sp-cwt-ks  &  3   &   \\
%           &  sp-ar-rc   &  23          &   \\
%            \bottomrule
%  \end{tabularx}
% % {\fnote {\textbf{TO}}: Time out: 5400 seconds for 64x64 and 128x128, 10800 seconds for the rest.}
%  % {\fnote {\textbf{TO\textsuperscript{1}}}: Time-out after 5400 seconds.}
%  % {\fnote {\textbf{TO\textsuperscript{2}}}: Time-out after 10800 seconds.}
 
% \end{table}

\subsection{Meta Functions for RP-Rewriter}
\label{sec:meta-rules}

Meta functions for  RP-Rewriter may be implemented the same  way as regular
ACL2 meta functions  excluding the support for {\it  stobj}s.  When proving
the correctness of meta functions, users may not use an arbitrary evaluator
but  have to  use our  evaluator  {\pcr rp-evlt},  which we  used to  prove
RP-Rewriter correct. Since {\pcr rp-evlt} might not recognize the functions
that users reason  about in their meta functions, we  implement a mechanism
to help this evaluator recognize new functions.  We adapt a system from FGL
libraries from  ACL2 Community books  (/centaur/fgl), and used  the utility
{\pcr def-formula-checks}, which creates an executable {\it formula-checks}
function with  a single argument,  state.  Execution of  the formula-checks
function  should  always evaluate  to  {\pcr  t}  unless some  function  is
redefined in ACL2.  This utility also  creates some rewrite rules for {\pcr
  rp-evlt} to  know how to evaluate  given new functions.  Suppose  that we
want to reason about a newly  defined function {\pcr (foo arg1 arg2)}, then
the generated rewrite rule for {\pcr foo} would have the following form:
\begin{verbnobox}[\fontfamily{pcr}\selectfont\fontsize{10}{12}\selectfont]
  (defthm rp-evlt-of-foo-when-example-formula-checks
    (implies (and (example-formula-checks state) 
                  (rp-evl-meta-extract-global-facts))
             (equal (rp-evlt (list 'foo arg1 arg2) a)
                    (foo (rp-evlt arg1 a)
                         (rp-evlt arg2 a)))))
\end{verbnobox} 
Having  created a  formula-checks  function {\pcr  example-formula-checks},
users  can prove  the  final  correctness theorem  for  evaluation of  meta
functions:
\begin{verbnobox}[\fontfamily{pcr}\selectfont\fontsize{10}{12}\selectfont]
  (defthm example-meta-fnc-correct
    (implies (and (example-formula-checks state) 
                  (rp-evl-meta-extract-global-facts)
                  (valid-sc term a)
                  (rp-termp term))
             (equal (rp-evlt (example-meta-fnc term) a)
                    (rp-evlt term a)))) 
\end{verbnobox}
Before  running  meta  functions,   we  run  the  generated  formula-checks
functions to relieve this extra hypothesis.

Functions  {\pcr valid-sc}  and {\pcr  rp-termp} check  the correctness  of
side-conditions  and  syntactic  coherence   of  terms,  respectively.   In
addition to term  evaluation with {\pcr rp-evlt}, users also  have to prove
another lemma stating that returned terms from meta functions satisfy {\pcr
  valid-sc}.  On the  other hand, proving that meta  functions return {\pcr
  rp-termp} is optional but recommended.  If  that proof is not provided by
the user, RP-Rewriter will run {\pcr  rp-termp} for all the returned terms,
which may slow down the proof significantly.  A more detailed discussion of
{\pcr rp-termp} and {\pcr valid-sc} is given in Section~\ref{sec:proof}.

% The function {\pcr rp-termp}
% defines the syntax of our terms. The innermost elements should either
% be quoted or a non-nil atom. If a term is an instance of {\pcr
%   falist}, then its arguments should be matching association lists as
% discussed in Section~\ref{sec:fast-alist}. If it is an instance of
% {\pcr rp}, then it should have two arguments, first of which should
% always be a quoted symbol. The term should never be a lambda
% expression. It is because we had difficulty proving the correctness
% with side-condition feature for lambda expressions.

A meta function may or may not  return {\it dont-rw} structure. If the user
chooses to generate {\it dont-rw}, then the return signature of the
meta function should be {\pcr (mv  term dont-rw)}.  It should be noted that
{\it dont-rw} structure is not the  only way to prevent redundant rewriting
of the terms returned from meta functions.  Users might also wrap the terms
that should  not be rewritten  (upon returning  from a meta  function) with
{\pcr hide}.  With a rule rewriting  {\pcr (hide x)} to {\pcr x}, instances
of {\pcr  hide} will  be removed  and rewriting on  selected terms  will be
avoided.  This alternative  method can also be used for  meta functions for
ACL2's rewriter.  In RP-Rewriter, users may choose either method to control
which terms should  be rewritten upon returning from  their meta functions.
Depending on  the application, using the  {\pcr hide} method might  be more
costly  than  {\it dont-rw}  because  creating  {\pcr hide}  instances  and
applying  a  rewrite rule  to  remove  them can  cause  more  memory to  be
allocated than  creating a {\it  dont-rw}.  This  memory gain might  not be
significant (for our multiplier proofs, we observe to use around 2-3\% less
memory);  however,   we  find   the  {\it  dont-rw}   method  to   be  more
convenient. Inserting numerous  {\pcr hide} instances inside  terms in meta
functions makes them  less readable.  Deciding later which  terms should or
should not  be rewritten  with the {\it  dont-rw} structure  facilitates an
easier development process.

\section{Applications}
\label{sec:uses}

RP-Rewriter has  been utilized  for various  applications that  may involve
dealing with very large terms. In this section, we briefly describe the use
cases for  multiplier design verification,  where RP-Rewriter is used  as a
clause processor,  and simplification of SV  expressions, where RP-Rewriter
is used as a logic-mode function.

\subsection{Multiplier Proofs}
\label{sec:mult-proofs}

We  use RP-rewriter  to implement  our verification  method for  multiplier
designs, which is based completely  on term rewriting \cite{mult-paper}. We
make  extensive   use  of  performance  benefits   of  side-conditions  and
fast-alist support.  We have tested the rewriter for very large multipliers
(i.e., 1024x1024),  whose proofs  create terms with  millions of  nodes but
finish in less than 10 minutes.

Our  multiplier verification  method is  based on  rewriting one-bit  adder
modules (e.g.,  full/half adders) in terms  of {\pcr mod} and  {\pcr floor}
functions  with rewrite  rules that  have  hypotheses that  all the  inputs
satisfy {\pcr  bitp}.  After  these rewrites, {\pcr  mod} and  {\pcr floor}
functions are simplified with some  other rewrite rules and meta functions,
and  the  circuit semantics  are  rewritten  to the  design  specification.
During this  arithmetic simplification,  the terms  change their  shapes so
much  that the  hypotheses in  our rewrite  rules become  too difficult  to
relieve similarly  to the scenario given  in Example \ref{d-evenp-example}.
Therefore we  attach side-conditions  to terms and  eliminate the  need for
backchaining  when   relieving  hypotheses  that  may   otherwise  be  very
expensive.

We  have  two  libraries  implementing  the  same  multiplier  verification
algorithm with different implementation techniques.  The first one is a mix
of  rewrite rules  and meta  functions, whereas  the second  implements the
majority of the  algorithm with a meta function for  better time and memory
performance.     They   are    published    in    ACL2   Community    books
(/books/projects/rp-rewriter/lib) (see :DOC  rp::multiplier-proofs and :DOC
rp::multiplier-proofs-2).   We do  not  have working  libraries for  ACL2's
built-in rewriter since these proofs depend on the side-conditions feature.

\subsection{SVEX Simplify}
\label{sec:svex-simplify}

A Symbolic  Vector Expression  (SVEX) is  a core data  type defined  with a
fixed  set of  functions to  describe  the behavior  of translated  Verilog
modules (see  :DOC sv::svex). In some  cases, these expressions can  be too
large and complex, and may need  to be simplified.  We implement a flexible
program with  RP-Rewriter to  take these expressions  saved as  a constant,
rewrite and  simplify them,  and return an  equivalent expression.  We call
this program {\pcr svex-simplify} (see :DOC svl::svex-simplify).

We convert  {\it SVEX}es  to regular ACL2  expressions using  its evaluator
{\pcr svex-eval}  with our  rewriter. The resulting  terms are  composed of
{\it 4vec}  functions, and  they are rewritten  and simplified  with proved
rewrite  rules.   Then the  resulting  terms  are  converted back  to  {\it
  SVEX}es. The {\pcr svex-simplify}  function with a simplification library
of    rewrite   rules    are    available   in    ACL2   Community    books
(/books/centaur/svl/). In this  system, we use the  main rewriter function,
{\pcr  rp-rw}, as  a  logic-mode  function instead  of  a clause  processor
because we  do not try  to prove  a conjecture but  take in a  constant and
return another.  Users may change the simplification method by disabling or
adding rewrite  rules to the system.   All the functions are  in logic mode
and guard-verified.

\section{Supported Rewrite Rules}
\label{sec:support}

RP-Rewriter supports only some of the rewrite rules. If a rule of any other
class, such as type-prescription, is  supplied to RP-rewriter, then it would
be treated  as a rewrite  rule. A rewrite  rule should match  the following
criteria:
\begin{itemize}
\item Hypotheses, LHS and RHS should satisfy {\pcr rp-termp}, which defines
  the syntax of our terms (see Section~\ref{sec:proof}).
\item Hypotheses,  LHS and RHS cannot  contain an instance of  {\pcr rp} or
  {\pcr  falist},   which  can   be  introduced   only  internally   or  by
  meta-functions.
\item Hypotheses and  RHS cannot contain a variable that  is not present in
  LHS (i.e., free variables and {\pcr bind-free} are not allowed)
\item LHS  cannot have an  instance of {\pcr  if} due to  some difficulties
  observed   during   verification   of    RP-Rewriter   with   regard   to
  side-conditions.
\end{itemize}

If a rewritten term has an instance of the form {\pcr (if test then else)},
the context (i.e., known facts) is expanded with {\pcr test} and {\pcr (not
  test)} when  the rewriter dives into  {\pcr then} and {\pcr  else} terms,
respectively. The side-conditions  under {\pcr then} and  {\pcr else} might
be valid with these expanded contexts.  When terms with side-conditions are
unified with the LHS of a rewrite rule, we add those side-conditions to the
context as well.   Keeping these in mind, consider that  we want to rewrite
this term:
\begin{verbnobox}[\fontfamily{pcr}\selectfont\fontsize{10}{12}\selectfont]
  (foo (if (p2 x) (rp 'p1 x) y) x)
\end{verbnobox}
\noindent
Only one of  the {\pcr x} instances has a  side-condition, and adding {\pcr
  (p1 x)}  to the context  would be unsound because  it might be  true only
when {\pcr (p2 x)}  is true.  We would have to  develop a special mechanism
for such cases.   Even though it is possible, we  decided not to complicate
our rewriting mechanism  and verification of RP-Rewriter. We  do not expect
to see many rewrite rules of this  form, and may start supporting it if the
need rises.

It should  be noted that  conjectures and  rewrite rules that  contain {\it
  lambda} expressions  are {\it beta-reduced} before  rewriting starts.  It
is due to  some difficulties experienced when  verifying the side-condition
feature (they present a completely different  class of terms, which is very
difficult to integrate), and we leave the support for lambda expressions as
a  future  work. For  cases  where  users  have  to maintain  {\it  lambda}
structure  in rewrite  rules to  prevent repeated  rewriting of  terms, our
utility {\it defthm-lambda}  can be used while proving the  rewrite rule in
ACL2.   This   utility  creates   different  functions  for   every  lambda
expression, and  divides the RHS of  the rule into multiple  rewrite rules.
This removes all the immediate lambda expressions from RHS but maintain the
same  functionality.   Example~\ref{defthm-lambda-example} shows  how  this
utility works.

\begin{example}
  \label{defthm-lambda-example}
  The  defthm-lambda  utility  replacing  a  rewrite  rule  with  a  lambda
  expression  on its  RHS with  an equivalent  rewrite rule  without lambda
  expressions.
\begin{verbnobox}[\fontfamily{pcr}\selectfont\fontsize{9}{12}\selectfont]
  (defthm-lambda foo-redef
    (implies (p x)
             (equal (foo x)
                    (let* ((a (f1 x))
                           (b (f2 x)))
                      (f4 a a b)))))
                           
  ;; The above event is translated into this:
  (encapsulate
    (((foo-redef_lambda-fnc_1 * *) => *)
     ((foo-redef_lambda-fnc_0 * *) => *))
    (local (defun-nx foo-redef_lambda-fnc_1 (b a)
             (f4 a a b)))
    (local (defun-nx foo-redef_lambda-fnc_0 (a x)
             (foo-redef_lambda-fnc_1 (f2 x) a)))
    
    (def-rp-rule foo-redef_lambda-opener
      (and (equal (foo-redef_lambda-fnc_1 b a)
                  (f4 a a b))
           (equal (foo-redef_lambda-fnc_0 a x)
                  (foo-redef_lambda-fnc_1 (f2 x) a))))

    (def-rp-rule foo-redef
      (implies (p x)
               (equal (foo x)
                      (foo-redef_lambda-fnc_0 (f1 x) x)))))
\end{verbnobox}
\end{example}

\section{Verification of RP-Rewriter}
\label{sec:proof}

RP-Rewriter  is  a verified  clause  processor.   We accomplished  this  by
creating a  proof scheme  with various  invariants throughout  our rewriter
functions.  These invariants are:
\begin{itemize}
\item Evaluation of terms with our evaluator {\pcr (rp-evlt term a)} should
  remain the same.
\item Evaluation for validity of  side-conditions with {\pcr (valid-sc term
    a)} should always return {\pcr t} .
\item The syntax of terms should satisfy {\pcr rp-termp}.
\end{itemize}
\noindent
In this section, we describe these invariants, and how meta-extract is used
to prove RP-Rewriter correct.

\subsection{Evaluation of Terms}

Every verified  meta function and  clause processor must be  proved correct
with  an evaluator  generated by  {\pcr defevaluator},  which is  a generic
utility  distributed with  ACL2. RP-Rewriter  is not  an exception,  and we
created an evaluator {\pcr rp-evl} using {\pcr defevaluator}, and we use an
extension of  that evaluator,  namely {\pcr  rp-evlt} that  processes terms
before evaluating.  For  every function that changes a term,  we prove that
the returned term evaluates to the same value as the input.

{\pcr  (rp-evlt term  a)} is  equivalent to  {\pcr (rp-evl  (rp-trans term)
  a)}. The terms we rewrite are  not regular translated ACL2 terms, but we
allow a sequence of  {\pcr cons} terms ending with {\pcr  nil} to be stored
and rewritten as {\pcr list} instances.   For example, the term {\pcr (cons
  a (cons  b (cons c  'nil)))} may be  stored as {\pcr  (list a b  c)}. The
function {\pcr  rp-trans} translates  the instances of  {\pcr list}  to its
equivalent  {\pcr cons}  form. This  is an  experimental feature,  and such
untranslated terms can only be created  and modified in meta functions. The
purpose of  this feature  is to  reduce memory  allocation when  very large
lists are present.

\subsection{Evaluation of Side-conditions}

The   regular  evaluator   is   expectedly  not   capable  of   recognizing
side-conditions  in our  terms. Therefore,  we create  a special  evaluator
{\pcr (valid-sc term a)} that  validates the correctness of side-conditions
that might  be present in terms.   Whenever there is a  proof of evaluation
with  {\pcr rp-evlt}  for  a function,  we  also have  a  proof with  {\pcr
  valid-sc}.  These proofs are in the following form:
\begin{verbnobox}[\fontfamily{pcr}\selectfont\fontsize{10}{12}\selectfont]
  (defthm some-rp-rw-fnc-valid-sc
    (implies (and (valid-sc term a)
                  <other-hypotheses>)
             (valid-sc (some-rp-rw-fnc term <other-args>) a))) 
\end{verbnobox}

{\pcr valid-sc}  is mutually-recursive  with {\pcr  valid-sc-subterms}, and
they traverse a term and  its subterms for side-conditions. {\pcr (valid-sc
  term a)} returns {\pcr  t} when {\pcr term} is an atom  or quoted.  If it
is matches the form {\pcr (if test then else)}, then it evaluates to:
\begin{verbnobox}[\fontfamily{pcr}\selectfont\fontsize{10}{12}\selectfont]
  (and (valid-sc test a)
       (if (rp-evlt test a)
           (valid-sc then a)
         (valid-sc else a))) 
\end{verbnobox}
\noindent
This  is because  we  test the  correctness under  the  current context  as
updated by such {\pcr if} calls. If  {\pcr term} is an instance of the form
{\pcr (rp 'prop x)}, then it evaluates:
\begin{verbnobox}[\fontfamily{pcr}\selectfont\fontsize{10}{12}\selectfont]
  (and (rp-evlt `(prop ,x) a) 
       (valid-sc x a))
\end{verbnobox}
\noindent
For   all  the   other   function  calls,   {\pcr   valid-sc}  runs   {\pcr
  valid-sc-subterms} on the arguments.  That  function in return runs {\pcr
  valid-sc} on each argument.

\subsection{Syntax of Terms}

Similar to {\pcr pseudo-termp}, we have {\pcr rp-termp} that set a standard
and rules for  the syntax of our  terms. We prove a  similar property about
{\pcr rp-termp}  for each of  our functions as  we do for  {\pcr valid-sc}.
{\pcr (rp-termp  term)} evaluates  to {\pcr t}  when {\pcr  term} satisfies
these conditions:
\begin{itemize}
\item Innermost elements should each either be a non-nil symbol or quoted.
\item Calls for {\pcr rp} should have two arguments, and the first argument
  should be a quoted non-nil symbol.
\item  Calls for  {\pcr falist}  should have  two arguments  and the  first
  argument  should be  a shadowing  fast-alist  of the  second argument  as
  described in  Section~\ref{sec:fast-alist}. This  is how we  maintain the
  invariant for the fast-alists feature.
\item Function names  must be a symbol and non-nil.  In other words, lambda
  expressions are not allowed.
\end{itemize}

Guaranteeing  that terms  retain this  syntax can  help prove  other lemmas
correct because it defines the shape  of terms and helps avoid dealing with
unexpected cases.

\subsection{Meta-extract}

Meta-extract  extends   the  capabilities  of  meta   functions  or  clause
processors to retrieve more facts from the {\tt state} \cite{meta-extract}.
The utility {\pcr def-meta-extract}  (see :DOC def-meta-extract), creates a
compact  macro for  evaluators that  can be  added to  the hypotheses  when
proving a meta-function/clause processor  correct.  For our evaluator {\pcr
  rp-evl}, this utility  creates {\pcr (rp-evl-meta-extract-global-facts)}.
When  we   add  this   call  to   our  hypotheses,   we  can   trust  {\pcr
  magic-ev-fncall}   (used   for    running   executable-counterparts   and
meta-functions added to RP-Rewriter),$\ $ {\pcr meta-extract-formula} (used
for extracting  rewrite/definition rules from ACL2)  and other meta-extract
functions to  return correct  values.  Working  with meta-extract  has been
fairly  easy,  and  it  did   not  complicate  our  proof  procedure  after
integration.

After establishing the correct  invariants and integrating meta-extract, we
could achieve a versatile and low-maintenance proof scheme that can hold up
well when changes are made or new features are added to the rewriter.

\section{Conclusion}
\label{sec:conclusion}

In this  paper, we have  introduced a rewriter customized  specifically for
conjectures  that may  expand into  very large  terms. We  have proved  the
correctness of this rewriter and saved it as a verified clause-processor in
ACL2. RP-Rewriter has the distinctive capability to retain properties about
terms  that  we  call   side-conditions.   With  other  features  including
fast-alist  support  and the  {\it  dont-rw}  structure, we  have  utilized
RP-Rewriter  for  applications  that  involved very  large  terms  such  as
multiplier design  verification and SVEX simplification.  Besides these use
cases, users might  also use rp-rewriter as a clause  processor on its own,
or as a part of their meta functions in order to perform term rewriting.

\section*{Acknowledgments}
\label{sec:ack}

We would  like to  specially thank Matt  Kaufmann for  detailed discussions
that help build the foundations of RP-Rewriter, J Strother Moore and Warren
A. Hunt,  Jr. for feedback, and  Sol Swords for his  suggestions to improve
the system.   This material is based  upon work supported in  part by DARPA
under Contract No. FA8650- 17-1-7704.

\nocite{*}
\bibliographystyle{eptcs}
\bibliography{rp-paper}
\end{document}